\documentclass[twocolumn,nofootinbib]{revtex4}
\usepackage[bookmarks = true, pdfpagemode = None, pdfstartview = FitH, colorlinks = true, urlcolor = blue]{hyperref}

\input{Qcircuit.tex}

\begin{document}

\title{Q-circuit Tutorial}

\author{Bryan Eastin, Steven T. Flammia}
\affiliation{Department of Physics and Astronomy, University of New
Mexico, Albuquerque, New Mexico 87131--1156, USA}

\begin{abstract}Q-circuit is a list of macros that greatly simplifies the construction of 
quantum circuit diagrams (QCDs) in \LaTeX \ with the help of the \Xy-pic 
package.  This tutorial should help the reader acquire 
the skill to render arbitrary QCDs in a matter of minutes.  Q-circuit is available for free\footnote{The Q-circuit package is distributed under the GNU public license.} at \href{http://info.phys.unm.edu/Qcircuit/}{http://info.phys.unm.edu/Qcircuit/}.
\end{abstract}

\maketitle

\section{Introduction}
\setcounter{footnote}{1}
Ever tried to use \LaTeX\ to typeset something like this?
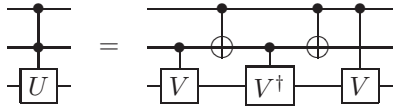
\[
\Qcircuit @C=.5em @R=0em @!R {
& \ctrl{1} & \qw & & & \qw & \ctrl{1} & \qw & \ctrl{1} & \ctrl{2} & \qw\\
& \ctrl{1} & \qw & \push{\rule{.3em}{0em}=\rule{.3em}{0em}} & & \ctrl{1} & \targ & \ctrl{1} & \targ & \qw & \qw\\
& \gate{U} & \qw & & & \gate{V} & \qw & \gate{V^\dag} & \qw & \gate{V} & \qw
}
\]
\noindent Or maybe this?
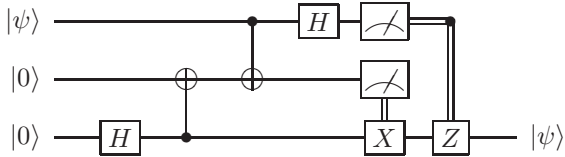
\[
\Qcircuit @C=.7em @R=.4em @! {
\lstick{\ket{\psi}} & \qw & \qw & \ctrl{1} & \gate{H} & \meter & \control \cw\\
\lstick{\ket{0}} & \qw & \targ & \targ & \qw & \meter & \cwx\\
\lstick{\ket{0}} & \gate{H} & \ctrl{-1} & \qw & \qw & \gate{X} \cwx & \gate{Z} \cwx & \rstick{\ket{\psi}} \qw
}
\]
\noindent Or how about\footnote{Code for these circuits is given in Appendix \ref{S:code}.}
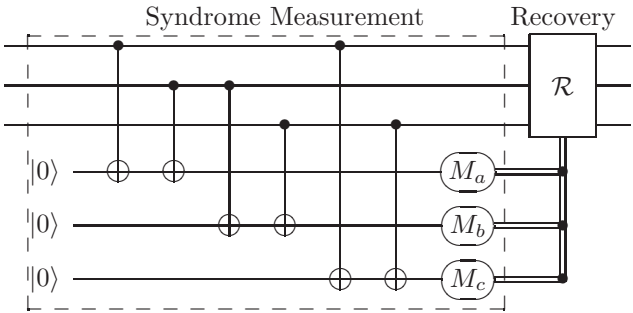
\[
\Qcircuit @C=1.3em @R=.6em {
  & & & & & & \mbox{Syndrome Measurement} & & & &
    \mbox{Recovery}\\
  & \qw & \qw & \ctrl{3} & \qw & \qw & \qw &
    \ctrl{5} & \qw & \qw &
    \multigate{2}{\ \mathcal{R}\ } & \qw\\
  & \qw & \qw & \qw & \ctrl{2} & \ctrl{3} & \qw &
    \qw & \qw & \qw & \ghost{\ \mathcal{R}\ } \qw &
    \qw\\
  & \qw & \qw & \qw & \qw & \qw & \ctrl{2} & \qw &
    \ctrl{3} & \qw & \ghost{\ \mathcal{R}\ } \qw &
    \qw\\
  & & \lstick{\ket{0}} & \targ \qw & \targ \qw &
    \qw & \qw & \qw & \qw & \measure{M_a} &
    \control \cw \cwx\\
  & & \lstick{\ket{0}} & \qw & \qw & \targ \qw &
    \targ \qw & \qw & \qw & \measure{M_b} &
    \control \cw \cwx\\
  & & \lstick{\ket{0}} & \qw & \qw & \qw & \qw &
    \targ \qw & \targ \qw & \measure{M_c}
    \gategroup{2}{2}{7}{10}{.8em}{--} &
    \control \cw \cwx
}
\]

Typesetting quantum circuit diagrams using standard \LaTeX\ graphics packages is a difficult and time consuming business.  Q-circuit is a high level macro package designed to change that.  With Q-circuit, drawing quantum circuit diagrams is as easy as constructing an array.  In a matter of minutes you can learn the basic syntax and start producing circuits of your own.

This tutorial teaches you to use Q-circuit from the ground up.  Many readers will find that they've learned everything they need to know by the end of \S\ref{S:basics}, but plenty of material is included for those that wish to typeset more complicated circuits.

% Q-circuit is free software*******Insert stuff about the GNU public license here*******.  If you use Q-circuit to render any quantum circuit diagrams in a document available to the public, please acknowledge that Q-circuit was used and, if appropriate, reference the website.  Please email me with suggestions and improvements.

\section{Getting Started}

To install Q-circuit, place the file \verb=Qcircuit.tex= somewhere your \TeX\ distribution can find it and run the appropriate command to update your \TeX\ tree.  To use it, place the command
{\small \begin{verbatim}\input{Qcircuit}\end{verbatim}}
\noindent in the preamble of your document.  \verb=Qcircuit.tex= loads the \verb=amsmath= and \verb=xy= packages and implements a set of circuit commands.  If need be, you can obtain the necessary packages at \href{http://www.ctan.org/}{http://www.ctan.org/}.

\section{Simple Quantum Circuits\label{S:basics}}

To begin, suppose the reader would like to typeset the following 
simple circuit:
\[ \Qcircuit @C=1em @R=.7em {
      & \gate{X} & \qw
}\]

This was typeset using
{\small \begin{verbatim}\Qcircuit @C=1em @R=.7em {
      & \gate{X} & \qw
}\end{verbatim}}

The command \verb=\Qcircuit= is simply a disguised \verb=\xymatrix= command 
with a default parameter set.  For readers unfamiliar with the \verb=xymatrix= environment, it suffices 
to know that it behaves more or less like the \verb=array= environment. 
That is, new columns are denoted by \verb=&= and new rows by \verb=\\=, 
as in the following example:
\[ \Qcircuit @C=1.4em @R=1.2em {
     a & i \\
     1 & x
} \]
which was typeset using
{\small \begin{verbatim}\Qcircuit @C=1.4em @R=1.2em {
     a & i \\
     1 & x
} \end{verbatim}}
The parameters \verb?@C=1.4em? and \verb?@R=1.2em? that appear after \verb=\Qcircuit= specify the spacing between the columns and the rows of the circuit, respectively.  They may take any length as an argument.  Additional parameters are discussed in \S\ref{S:spaces}.

\subsection{Wires and gates}

The command \verb=\qw= draws a wire between two columns of 
a QCD. The command derives its name from an abbreviation of `quantum wire'.  
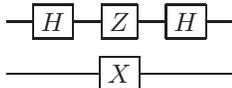
\[ \Qcircuit @C=1em @R=.7em {
   & \gate{H} & \gate{Z} & \gate{H} & \qw \\
   & \qw & \gate{X} & \qw & \qw
} \]

The diagram above was drawn using 
{\small \begin{verbatim}\Qcircuit @C=1em @R=.7em {
   & \gate{H} & \gate{Z} & \gate{H} & \qw \\
   & \qw & \gate{X} & \qw & \qw
}\end{verbatim}}
\noindent Note that \verb=\qw= is used to connect a wire {\it towards the left}.

The \verb=\gate= command draws the argument of the function inside a 
framed box and extends a wire {\it back to the previous column}.  When using the 
\verb=\gate= and \verb=\qw= commands, make sure there is another column 
entry to the left of the current column entry in your QCD, otherwise the wire will not 
connect to anything (and you'll get an error), as in the following example code:

{\small \begin{verbatim} (**Wrong!**)
\Qcircuit @C=1em @R=.7em {
     \gate{U} & \qw \\
     \gate{U^\dag} & \qw
} \end{verbatim}}

The proper way to render this circuit would be to include space for the incoming
wires at the beginning by inserting the \verb=&= character at the start of each new line:
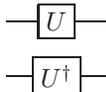
\[ \Qcircuit @C=1em @R=.7em {
     & \gate{U} & \qw \\
     & \gate{U^\dag} & \qw
} \]

{\small \begin{verbatim}\[ \Qcircuit @C=1em @R=.7em {
     & \gate{U} & \qw \\
     & \gate{U^\dag} & \qw
} \]\end{verbatim}}

The only difference between these two codes is that the correct code has an ampersand (\verb=&=) at the start of each new line.

\subsection{CNOT and other controlled single qubit gates \label{S:CNOT}}

With just these few commands, one can already render a circuit with an arbitrary number of wires and single qubit gates.  In this section, we'll learn how to draw CNOT 
gates and controlled single qubit gates with an arbitrary number 
of controls.  

A simple circuit with two CNOT gates in it is
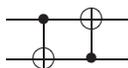
\[ \Qcircuit @C=1em @R=.7em {
     & \ctrl{1} & \targ & \qw \\
     & \targ & \ctrl{-1} & \qw
}\]
which was typeset by
{\small \begin{verbatim}\Qcircuit @C=1em @R=.7em {
     & \ctrl{1} & \targ & \qw \\
     & \targ & \ctrl{-1} & \qw
}\end{verbatim}}
In this circuit, the command \verb=\targ= draws the target gate on the 
wire, and the \verb=\ctrl{#1}= puts a bullet down, and connects to the 
target which is \verb=#1= array elements \textit{below} the control.
Hence, to connect the second CNOT gate properly, we used -1.

A more complicated circuit with multiple controls and arbitrary gates
might look like
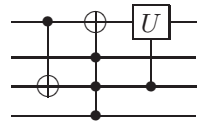
\[ \Qcircuit @C=1em @R=.7em {
   & \ctrl{2} & \targ & \gate{U} & \qw \\
   & \qw & \ctrl{-1} & \qw & \qw \\
   & \targ & \ctrl{-1} & \ctrl{-2} & \qw \\
   & \qw & \ctrl{-1} & \qw & \qw 
}\]
which was drawn using
{\small \begin{verbatim}\Qcircuit @C=1em @R=.7em {
   & \ctrl{2} & \targ & \gate{U} & \qw \\
   & \qw & \ctrl{-1} & \qw & \qw \\
   & \targ & \ctrl{-1} & \ctrl{-2} & \qw \\
   & \qw & \ctrl{-1} & \qw & \qw 
}\end{verbatim}}

In the first gate, the control bit connects to the 
target on wire 3.  In the second gate, each control connects to the object 
directly above it.  Finally, the third gate is an example of how to do 
controls on arbitrary gates; simply place the desired gate where you would
normally put a target.

\subsection{Vertical wires}

Suppose we want to typeset the following circuit:
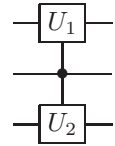
\[ \Qcircuit @C=1em @R=1.2em {
     & \gate{U_1} & \qw \\
     & \ctrl{1} \qwx & \qw \\
     & \gate{U_2} & \qw \\
}\]
so that the middle control has to connect to more than one gate.  The way
to accomplish this is with the \verb=\qwx= command.  The command
\verb=\qwx[#1]= takes an optional input, \verb=#1=, and connects from 
the current position to a position \verb=#1= entries {\it below}
the current position.  The default argument is -1.  Thus, one way to
typeset the above diagram is with the following code:
{\small \begin{verbatim}\Qcircuit @C=1em @R=1.2em {
     & \gate{U_1} & \qw \\
     & \ctrl{-1} \qwx[1] & \qw \\
     & \gate{U_2} & \qw \\
}\end{verbatim}}
\noindent or, equivalently,
{\small \begin{verbatim}\Qcircuit @C=1em @R=1.2em {
     & \gate{U_1} & \qw \\
     & \ctrl{1} \qwx & \qw \\
     & \gate{U_2} & \qw \\
}\end{verbatim}}
\noindent which is what the author used.

Note that wire commands must not precede the gate command in an entry.
Also, remember that commands taking an optional 
argument use {\it square} braces rather than curly braces.

\subsection{Labelling input and output states \label{S:labels}}
The last element we need for simple circuits is the ability to add labels.  We'll look at input and output labels here, other kinds of labels are discussed in \S\ref{S:labels2}.

When labelling input and output qubits, one should use the \verb=\lstick= and \verb=\rstick= commands.  These commands ensure that the labels and the wires connecting to them line up correctly.  The \verb=\lstick= command is used for input labels (on the left of the diagram), and the \verb=\rstick= command is used for output labels (on the right of the diagram).  Placement rules are the same as those for gates with the exception that \verb=\lstick= and \verb=\rstick= can be inserted in the leftmost column of the array.  Here is an example circuit:
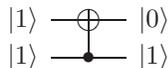
\[ \Qcircuit @C=1em @R=1em {
\lstick{\ket{1}} & \targ &  \rstick{\ket{0}} \qw \\
\lstick{\ket{1}} & \ctrl{-1} & \rstick{\ket{1}} \qw
}\]
typeset with
{\small \begin{verbatim}\Qcircuit @C=1em @R=1em {
 \lstick{\ket{1}} & \targ & \rstick{\ket{0}} \qw \\
 \lstick{\ket{1}} & \ctrl{-1} & \rstick{\ket{1}} \qw
}\end{verbatim}}

\section{More Complicated Circuits: Multiple Qubit Gates and Beyond}

So far, we have seen how to make arbitrary QCDs involving single qubit gates and controlled gates, including CNOT.  Since this is known to be universal for computation, we could just stop here!  Of course, many circuit diagrams use more complicated structures such as multi-qubit gates, measurements, classical wires, and swaps.  We will learn how to use Q-circuit to make all of these in this section.

\subsection{Multiple qubit gates \label{S:multigate}}

Let's look at an example, and then we'll explain the code. 
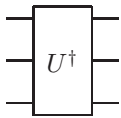
\[ \Qcircuit @C=1em @R=.7em {
     & \multigate{2}{U^\dag} & \qw \\
     & \ghost{U^\dag}& \qw \\
     & \ghost{U^\dag} & \qw 
}\]
The 3-qubit gate above was typeset with
{\small \begin{verbatim}\Qcircuit @C=1em @R=.7em {
     & \multigate{2}{U^\dag} & \qw \\
     & \ghost{U^\dag}& \qw \\
     & \ghost{U^\dag} & \qw 
}\end{verbatim}}
First let's go over the \verb=\multigate= command.  
\verb=\multigate{#1}{#2}= is a two argument gate that takes the 
\textit{depth} of the
gate for the first argument and the \textit{label} of the gate for the
second argument.  In the above example, \verb=#1= equals 2 because the 3-qubit gate
extends two rows below the position of \verb=\multigate=.  On the other two lines, 
the \verb=\ghost= command is used to get the spacing and connections right.  \verb=\ghost= behaves like an invisible gate that allows the quantum wires on either side of your multigate to connect correctly.

The generalization to an arbitrarily large gate is now obvious.  Let's look at a 6-qubit gate. The code
{\small 
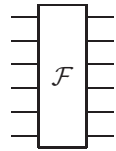
\begin{verbatim}\Qcircuit @C=1em @R=0em {
     & \multigate{5}{\mathcal{F}} & \qw \\
     & \ghost{\mathcal{F}} & \qw \\
     & \ghost{\mathcal{F}} & \qw \\
     & \ghost{\mathcal{F}} & \qw \\
     & \ghost{\mathcal{F}} & \qw \\
     & \ghost{\mathcal{F}} & \qw 
}\end{verbatim}}
\noindent yields
\[ \Qcircuit @C=1em @R=0em {
     & \multigate{5}{\mathcal{F}} & \qw \\
     & \ghost{\mathcal{F}} & \qw \\
     & \ghost{\mathcal{F}} & \qw \\
     & \ghost{\mathcal{F}} & \qw \\
     & \ghost{\mathcal{F}} & \qw \\
     & \ghost{\mathcal{F}} & \qw 
}\]
Thus, for every entry below the top, a \verb=\ghost= command
with the label for the gate is needed.  Strictly speaking, the name of the gate is not necessary inside the \verb=\ghost= command.  Since \verb=\ghost= is just an invisible place holder, anything with the same width as the label specified in multigate will work as well.  In practice, however, it is usually easiest to use the same argument.

Note that controls to multiple qubit gates work the same as for single
qubit gates, using \verb=\ctrl= and \verb=\qwx=.

\subsection{Measurements and classical bits}

Measurement gates are typeset just like ordinary gates, but they typically have some sort of decoration to indicate that measurement has occurred.  At present, Q-circuit supports the following single qubit measurement gates.
{\small \begin{center}
    \begin{tabular}{l | l | l} 
        \multicolumn{1}{c}{\itshape Example} & \multicolumn{1}{c}{\itshape Command} & \multicolumn{1}{c}{\itshape Example Code }\\ \hline 
        \Qcircuit @C=1em @R=.7em {& \meter}
            & \verb=\meter= & \verb=\meter=\\
        \Qcircuit @C=1em @R=.7em {& \measure{\mbox{Basis}}}
            & \verb=\measure = & \verb=\measure{\mbox{Basis}}=\\
        \Qcircuit @C=1em @R=.7em {& \measuretab{M_{ijk}}} \hspace{.5em}
            & \verb=\measuretab= & \verb=\measuretab{M_{ijk}}=\\
        \Qcircuit @C=1em @R=.7em {& \measureD{\chi}}
            & \verb=\measureD= & \verb=\measureD{\chi}=
    \end{tabular}
\end{center}}

Often we want to condition some gate on the output of a measurement.  One convenient way illustrate this is with the classical wire commands, \verb=\cw= and \verb=\cwx=.  The classical wire commands work exactly like the quantum wire commands, but they draw double instead of single lines.

Here is an example using measurement gates and classical wires and the corresponding code.
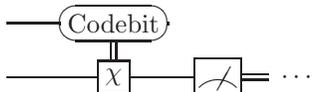
\[\Qcircuit @C=1em @R=.7em {
     & \qw & \measure{\mbox{Codebit}} \cwx[1] \\
     & \qw & \gate{\chi} & \meter & \rstick{\cdots} \cw
}\]
{\small \begin{verbatim}\Qcircuit @C=1em @R=.7em {
     & \qw & \measure{\mbox{Codebit}} \cwx[1] \\
     & \qw & \gate{\chi} & \meter &
        \rstick{\cdots} \cw
}\end{verbatim}}

Q-circuit also includes the commands \verb=\multimeasure= and \verb=\multimeasureD= for typesetting measurements on multiple qubits.  The syntax for these commands exactly parallels that of the \verb=\multigate= command (see \S\ref{S:multigate}).  An example is shown below.
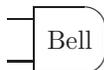
\[\Qcircuit @C=1em @R=.7em {
    & \multimeasureD{1}{\text{Bell}} \\
    & \ghost{\text{Bell}}
}\]
{\small \begin{verbatim}\Qcircuit @C=1em @R=.7em {
    & \multimeasureD{1}{\text{Bell}} \\
    & \ghost{\text{Bell}}
}\end{verbatim}}
 
\subsection{Non-gate inserts, forcing space, and swap \label{S:inserts}}

In addition to the gates defined by Q-circuit, standard \LaTeX\ can function as a gate if enclosed in curly brackets.  By default, inputs are assumed to have zero size, so no space will be made for the resulting object and any wires connecting to it will run straight to the object's middle.  Standard \LaTeX\ entries can serve as labels or wire decorations.

To force an object to take up space, you should use the \verb=\push= command.  \verb=\push= is most useful in conjunction with the \LaTeX\ command \verb=\rule=.  Together they can be used to construct various sorts of invisible props and struts.

Q-circuit implements a gate command called \verb=\qswap= that is equivalent to the text \verb={\times} \qw=.  The effect of \verb=\qswap= is to insert half of a swap gate (that is a $\times$) which can then be connected (using \verb=\qwx=) to another instance of \verb=\qswap= to create a swap gate.

Here is a circuit that shows how to construct swap, decorate wires, and use \verb=\push= to make an invisible prop.
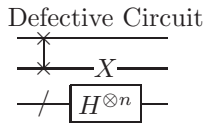
\[\Qcircuit @C=1em @R=.3em {
     &  & \mbox{Defective Circuit}\\
     & \qswap & \qw & \push{\rule{0em}{1em}} \qw \\
     & \qswap \qwx & \push{X} \qw & \qw \\
     & {/} \qw & \gate{H^{\otimes n}} & \qw
}\]
{\small \begin{verbatim}\Qcircuit @C=1em @R=.3em {
     &  & \mbox{Defective Circuit}\\
     & \qswap & \qw & \push{\rule{0em}{1em}} \qw \\
     & \qswap \qwx & \push{X} \qw & \qw \\
     & {/} \qw & \gate{H^{\otimes n}} & \qw
}\end{verbatim}}

\subsection{How to control anything}

Controlled-Z gates, wires with bends, and gates that control-on-zero can all be made using the extended family of control commands.  The complete family of control commands is \verb=\ctrl=, \verb=\crtlo=, \verb=\control=, and \verb=\controlo=.

\verb=\ctrlo= is identical to the \verb=\ctrl= command (see \S\ref{S:CNOT}) except that it draws an open bullet (indicating control-on-zero).  Both commands place a wire to the left and take one argument indicating which wire to connect to.

The commands \verb=\control= and \verb=\controlo= are isolated controls; they don't automatically connect to anything.  Isolated controls allow you to decide exactly what connections are made to your control operator, which makes them very useful for working with classical wires and rendering things like the controlled-Z.

Here is an example circuit using various controls.  
\[ \Qcircuit @C=1em @R=.7em {
    & \ctrl{2} & \ctrlo{1} & \ctrl{1} & \qw & \multigate{1}{U} & \qw \\
    & \qw & \targ & \ctrlo{2} \qw & \ctrl{1} & \ghost{U} & \qw\\
    & \control \qw & \ctrl{1} & \qw & \meter & \controlo \cw \cwx &\\
    & \qw & \control \qw & \gate{H} & \meter & \control \cw \cwx
}\]

{\small \begin{verbatim}\Qcircuit @C=1em @R=.7em {
    & \ctrl{2} & \ctrlo{1} & \ctrl{1} 
        & \qw & \multigate{1}{U} & \qw \\
    & \qw & \targ & \ctrlo{2} \qw 
        & \ctrl{1} & \ghost{U} & \qw \\
    & \control \qw & \ctrl{1} & \qw 
        & \meter & \controlo \cw \cwx \\
    & \qw & \control \qw & \gate{H} 
        & \meter & \control \cw \cwx
}\end{verbatim}}

Note that we, the authors, have used a pair of controls connected by a wire to denote the controlled-Z gate.  This isn't standard notation, but we feel it is a logically consistent and concise notation, and it illustrates nicely the symmetry of the controlled-Z gate.  We hope to encourage the readers to adopt this notation in their own QCDs.

\section{Bells and Whistles: Tweaking Your Diagram to Perfection}

By now, the reader should be able to quickly and easily typeset almost any QCD.  Nonetheless, it may occasionally be desirable to decorate or modify a circuit in ways not yet discussed.  This section presents additional tricks, options, and commands for putting the final polish on your QCDs.

\subsection{Spacing\label{S:spaces}}

The Q-circuit parameters \verb+@R+ and \verb+@C+ were introduced in \S\ref{S:basics}; they are examples of a family of spacing parameters that can appear between the text \verb=\Qcircuit= and the opening curly brace.  A more complete list of available parameters is given in the table below.

{\small \begin{center}
    \begin{tabular}{l | l } 
        \multicolumn{1}{c}{\itshape Parameter} & \multicolumn{1}{c}{\itshape Effect }\\ \hline 
        \verb+@R=#1+ & Sets the spacing between rows to \verb=#1=.\\
        \verb+@C=#1+ & Sets the spacing between columns to \verb=#1=.\\
        \verb+@!R+ & \parbox[t]{6cm}{Sets all rows to the height of the tallest object in the circuit.}\\
        \verb+@!C+ & \parbox[t]{6cm}{Sets all columns to the width of the widest object in the circuit.}\\
        \verb+@!+ & \parbox[t]{6cm}{Sets all entries to the size of the largest object in the circuit.}
    \end{tabular}
\end{center}}

The \verb=@R= and \verb=@C= parameters adjust the separation between elements, allowing you to dictate the compactness of your QCD.  \verb=@!R=, \verb=@!C=, and \verb=@!= force the elements of your circuit to have uniform sizes, this helps to prevent bunching that may occur when a particular row or column contains many small elements. \verb=@!R= is particularly useful for forcing wires to be evenly spaced, as in the following example.
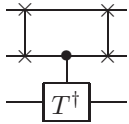
\[ \Qcircuit @C=.7em @R=.3em @!R {
    & \qswap & \qw & \qswap & \qw\\
    & \qswap \qwx & \ctrl{1} & \qswap \qwx & \qw \\
    & \qw & \gate{T^\dag} & \qw & \qw
}\]

{\small \begin{verbatim}\Qcircuit @C=.7em @R=.3em @!R {
    & \qswap & \qw & \qswap & \qw\\
    & \qswap \qwx & \ctrl{1} & \qswap \qwx & \qw \\
    & \qw & \gate{T^\dag} & \qw & \qw
}\end{verbatim}}

\subsection{Labelling \label{S:labels2}}

A label can be placed anywhere that a gate command might normally appear.  Unlike gates, however, Q-circuit treats labels as having zero size when determining the layout of a QCD.  This prevents large labels from bending your circuit out of whack, but it also means that labels can overlap with other components.

Normally an element whose size is set to zero is drawn centered on it's entry.  This is what happens when you insert text directly using curly brackets (see \S\ref{S:inserts}).  For most labelling, however, it is more useful to have one edge of the label fixed in the center of an entry.  For this reason Q-circuit provides a set of label commands, \verb=\lstick=, \verb=\rstick=, \verb=\ustick=, and \verb=\dstick=.  The stick commands each cause their contents to ``stick out" from the center of an entry in a different direction.  \verb=\lstick=, \verb=\rstick=, \verb=\ustick=, and \verb=\dstick= produce labels that project out to the left, right, top, and bottom respectively.

Proper usage of \verb=\lstick= and \verb=\rstick= was demonstrated in \S\ref{S:labels}, so the following example focuses on \verb=\ustick= and \verb=\dstick=.
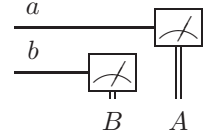
\[ \Qcircuit @C=.7em @R=.3em {
    & \ustick{a} \qw & \qw & \qw & \qw & \meter \\
    & \ustick{b} \qw & \qw & \qw & \meter \\
    & & & & \dstick{B} \cwx & \dstick{A} \cwx[-2]
}\]
\\
{\small \begin{verbatim}\Qcircuit @C=.7em @R=.3em {
    & \ustick{a} \qw & \qw & \qw & \qw & \meter \\
    & \ustick{b} \qw & \qw & \qw & \meter \\
    & & & & \dstick{B} \cwx & \dstick{A} \cwx[-2]
}\end{verbatim}}

\subsection{Grouping}

It is sometimes useful to box off sections of a circuit to indicate a subcircuit, as in the following example.
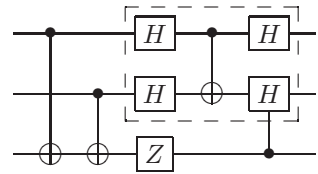
\[ \Qcircuit @C=1em @R=1em {
    & \ctrl{2} & \qw & \gate{H} & \ctrl{1} & \gate{H} & \qw \\
    & \qw & \ctrl{1} & \gate{H} & \targ & \gate{H} & \qw \\
    & \targ & \targ & \gate{Z} & \qw & \ctrl{-1} & \qw \gategroup{1}{4}{2}{6}{.7em}{--}
}\]
which was typeset using
{\small \begin{verbatim}\Qcircuit @C=1em @R=1em {
    & \ctrl{2} & \qw & \gate{H} & \ctrl{1} & 
        \gate{H} & \qw \\
    & \qw & \ctrl{1} & \gate{H} & \targ &
        \gate{H} & \qw \\
    & \targ & \targ & \gate{Z} & \qw & \ctrl{-1} &
        \qw \gategroup{1}{4}{2}{6}{.7em}{--} 
}\end{verbatim}}
The command that made the dashed box is in the last line of code and is called \verb=\gategroup=.  The \verb=\gategroup= command can be placed following any non-empty entry, but, for clarity, it is perhaps best to put it at the end.

Because it takes six arguments, \verb=\gategroup= looks intimidating, but it is actually relatively easy to use.  \verb=\gategroup{#1}{#2}{#3}{#4}{#5}{#6}= highlights the entries between rows \verb=#1= and \verb=#3= and columns \verb=#2= and \verb=#4= by adding a box or a bracket.  Argument \verb=#6= selects between various highlights, with the available options being:
\begin{center} {\small \verb=--  .  _\}  ^\}  \{  \}  _)  ^)  (  )=} \end{center}
These options produce a dashed box, a dotted box, a curly brace on the bottom, top, left, or right, and a normal brace on the bottom, top, left, or right.  Argument \verb=#5= is twice the spacing from the nearest gate to the box.

\verb=\gategroup= only checks that the gates at the four corners of the requested region are properly enclosed.  As a result, gates along the boundary that are bigger than the corner gates will tend to stick out.  This is especially unsightly when the corner entries are wires, though in that case the problem can be fixed by inserting an invisible prop of the desired height (see \S\ref{S:inserts}).

\section{Acknowledgments}

The authors would like to thank Aaron Smith, Joe Renes, and Andrew Silberfarb for useful discussions, ideas, and debugging.  Thanks to Carl Caves and Michael Nielsen for encouragement on this project.  An extra thanks to Michael Nielsen for suggesting some useful \LaTeX \ resources.

The development of Q-circuit was supported in part by the National Security Agency (NSA) and the Advanced Research and Developement Activity (ARDA) under the Army Research Office (ARO) contract numbers DAAD19-01-1-0648 and W911NF-04-1-0242.

\appendix
\section{Positioning Q-circuit diagrams in \LaTeX}
Q-circuit produces \TeX\ graphics objects.  In theory these objects should act like any symbol or character.  Thus, they can be placed in equation environments, arrays, and figures.  In practice there are a few, largely unexplained, complications.

One of these is vertical centering in a line of text.  To center the top line of a circuit, it is sufficient to invoke it in inline math mode using \verb=$=.  To center the entire circuit, place it inside an array.

Horizontal centering within figures is also problematic.  Typically this can be corrected by placing the \verb=\Qcircuit= command inside a \verb=\centerline= command, an \verb=\mbox= command, or an equation environment.  For some \LaTeX\ distributions the commands \verb=\leavevmode= and \verb=\centering= must be added to center a figure. 

Finally, circuits using large labels often appear a bit off center.  This is because labels are not included when calculating the size of a circuit.  The best solution is probably to add white space (see \S\ref{S:inserts}) until the labels all fit within the boundaries of the circuit.

\section{Bugs and Future Work}

\begin{enumerate}
\item Wires often end just short of curved surfaces.
\item \verb=\gategroup= needs to check all the boundary gates when determining the highlighted area.
\item Targets look poor when the font size is set to small.
\item It would be nice if the \verb=\ghost= command could read the argument of the \verb=\multigate= command automatically.
\item Larger issues of centering within \LaTeX\ need to be addressed.
\end{enumerate}

\section{Code for the Introduction\label{S:code}}
The first QCD depicts a way of decomposing doubly controlled unitaries.
It was typeset with
{\small \begin{verbatim}\Qcircuit @C=.5em @R=0em @!R {
  & \ctrl{1} & \qw & & & \qw & \ctrl{1} & \qw &
    \ctrl{1} & \ctrl{2} & \qw\\
  & \ctrl{1} & \qw & 
    \push{\rule{.3em}{0em}=\rule{.3em}{0em}} & &
    \ctrl{1} & \targ & \ctrl{1} & \targ & \qw &
    \qw\\
  & \gate{U} & \qw & & & \gate{V} & \qw &
    \gate{V^\dag} & \qw & \gate{V} & \qw
}
\end{verbatim}}

The second QCD depicts quantum teleportation and was typeset with
{\small \begin{verbatim}\Qcircuit @C=.7em @R=.4em @! {
  \lstick{\ket{\psi}} & \qw & \qw & \ctrl{1} &
    \gate{H} & \meter & \control \cw\\
  \lstick{\ket{0}} & \qw & \targ & \targ & \qw &
    \meter & \cwx\\
  \lstick{\ket{0}} & \gate{H} & \ctrl{-1} & \qw &
    \qw & \gate{X} \cwx & \gate{Z} \cwx &
    \rstick{\ket{\psi}} \qw
}
\end{verbatim}}

The third QCD depicts quantum error correction on the bit flip code.  It was typeset with
{\small \begin{verbatim}\Qcircuit @C=1.3em @R=.6em {
  & & & & & & \mbox{Syndrome Measurement} & & & &
    \mbox{Recovery}\\
  & \qw & \qw & \ctrl{3} & \qw & \qw & \qw &
    \ctrl{5} & \qw & \qw &
    \multigate{2}{\ \mathcal{R}\ } & \qw\\
  & \qw & \qw & \qw & \ctrl{2} & \ctrl{3} & \qw &
    \qw & \qw & \qw & \ghost{\ \mathcal{R}\ } \qw &
    \qw\\
  & \qw & \qw & \qw & \qw & \qw & \ctrl{2} & \qw &
    \ctrl{3} & \qw & \ghost{\ \mathcal{R}\ } \qw &
    \qw\\
  & & \lstick{\ket{0}} & \targ \qw & \targ \qw &
    \qw & \qw & \qw & \qw & \measure{M_a} &
    \control \cw \cwx\\
  & & \lstick{\ket{0}} & \qw & \qw & \targ \qw &
    \targ \qw & \qw & \qw & \measure{M_b} &
    \control \cw \cwx\\
  & & \lstick{\ket{0}} & \qw & \qw & \qw & \qw &
    \targ \qw & \targ \qw & \measure{M_c}
    \gategroup{2}{2}{7}{10}{.8em}{--} &
    \control \cw \cwx
}
\end{verbatim}}

\pagebreak

\section{Table of Commands}

The following table is grouped according to the effect of each command.\\

{\small \begin{center}
    \begin{tabular}{l | l } 
        \multicolumn{1}{c}{\itshape Subject} & \multicolumn{1}{c}{\itshape Command }\\ \hline 
        Loading Q-circuit \hspace{.5em} & \verb=\input{Qcircuit}= \\
        Making Circuits & \verb=\Qcircuit= \\
        Spacing & \parbox[t]{6cm}{\tt
                    @C=\#1 \\
                    @R=\#1 \\
                    @!R \\
                    @!C \\
                    @! \\
                    \char92 push\{\#1\}} \\
        Wires & \parbox[t]{6cm}{\tt
                    \char92 qw[\#1] \\
                    \char92 qwx[\#1] \\
                    \char92 cw[\#1] \\
                    \char92 cwx[\#1] }\\
        Gates & \parbox[t]{6cm}{\tt
                    \char92 gate\{\#1\} \\
                    \char92 targ \\
                    \char92 qswap \\
                    \char92 multigate\{\#1\}\{\#2\} \\
                    \char92 ghost\{\#1\} }\\
        Controls & \parbox[t]{6cm}{\tt
                    \char92 ctrl\{\#1\} \\
                    \char92 ctrlo\{\#1\} \\
                    \char92 control \\
                    \char92 controlo } \\
        Measurements & \parbox[t]{6cm}{\tt
                    \char92 meter \\
                    \char92 measure\{\#1\} \\
                    \char92 measureD\{\#1\} \\
                    \char92 measuretab\{\#1\} \\
                    \char92 multimeasure\{\#1\}\{\#2\} \\
                    \char92 multimeasureD\{\#1\}\{\#2\} } \\
        Labels & \parbox[t]{6cm}{\tt
                    \char92 lstick\{\#1\} \\
                    \char92 rstick\{\#1\} \\
                    \char92 ustick\{\#1\} \\
                    \char92 dstick\{\#1\} \\
                    \char92 bra\{\#1\} \\
                    \char92 ket\{\#1\} \\
                    \char92 gategroup\{\#1\}\{\#2\}\{\#3\}\{\#4\}\{\#5\}\{\#6\} }
    \end{tabular}
\end{center}}

\end{document}